# Stable products of laser-induced breakdown of aqueous colloidal solutions of nanoparticles

E.V. Barmina[1], S.V. Gudkov[1], A.V. Simakin[1], and G.A. Shafeev[1,2*]

[1]A.M. Prokhorov General Physics Institute of the Russian Academy of Sciences
[2]National Research Nuclear University MEPhI (Moscow Engineering Physics Institute), 31, Kashirskoye highway, 115409, Moscow, Russian Federation
* Corresponding author, e-mail: shafeev@kapella.gpi.ru

Abstract

The formation of stable products of water decomposition under laser exposure of aqueous colloidal solutions of nanoparticles is experimentally studied. Laser exposure of colloidal solutions leads to formation of $H_2$, $O_2$, and $H_2O_2$. The dependence of the yield of these products depends on the energy density of laser radiation inside the liquid and concentration of nanoparticles. The ratio $H_2/O_2$ depends on laser fluence and is shifted towards $H_2$. There are at least to sources of $H_2O_2$, namely, laser-induced breakdown plasma and ultrasound induced by laser pulses in the liquid. The formation of both $H_2$ and $O_2$ is tentatively assigned to direct dissociation of $H_2O$ molecules by electron impact from laser-induced plasma.

Introduction

Laser ablation of solids in liquids is a physical method of generation of large variety of nanoparticles. Typically, a solid target is placed into liquid, which is transparent for laser radiation. If the laser fluence on the target is enough for its melting, then this melt is dispersed into surrounding liquid as nanoparticles (NPs). Virtually any liquid is suitable for generation of NPs in this way, water being the most common of them. Organic solvents, such as alcohols, are also frequently used for preparation of colloidal solutions of different NPs in them. Laser ablation of solids in liquids is accompanied by formation of plasma plume above the target. Plasma is also observed under laser exposure of colloidal solutions of NPs. If laser intensity is high enough for NPs to reach temperatures of about $10^4$-$10^5$ K, some part of their atoms may be ionized [1]. Most of the liquids used for laser generation of NPs contain hydrogen or its isotopes. In conditions of laser-induced breakdown both the liquid and the material of NPs are affected resulting in chemical changes of their composition. In case of organic solvents, e.g., ethanol, this leads to deep pyrolysis of the liquid down to formation of elementary glassy carbon [2]. One should also expect the formation of gaseous products of liquid decomposition. Indeed, as it was recently shown, laser exposure of colloidal solution of Au nanoparticles in water is accompanied by emission of molecular $H_2$ [3]. There should be other stable products of $H_2O$ molecules decomposition, for example, molecular $O_2$. In this work we present new results concerning the formation of both molecular $O_2$ and $H_2$ simultaneously and independently measured under laser exposure of $H_2O$ containing small nanoparticles [4]. Different pathways of formation of these products are discussed.

Experimental technique

Experimental setup is described elsewhere [4]. We used technical water that contained about $10^{11}$ cm$^{-3}$ particles with average size of 2.5 nm (as determined with measuring disk centrifuge) and made mostly of $Fe_3O_4$ and $Fe(OH)_2$. Qualitatively the results remain the same using the purest H2O available (conductivity of 0.6 μSm) though the presence of nanoparticles leads to higher rate of gases production. The role of nanoparticles is to ignite the primary breakdown of the colloidal solution which then develops into microscopic breakdown channel of the liquid. Irradiation of NPs colloidal solutions diluted in necessary proportion by pure water was carried out using the radiation of the Nd:YAG laser at wavelength of 1064 nm and pulse duration of about 10 ns (FWHM). Laser radiation was focused inside the liquid by an F-Theta objective with focal distance of 90 mm. Laser beam was scanned across the window along circular trajectory about 8 mm in diameter at the velocity of 3000 mm/s by means of galvo mirror system. Laser exposure of colloids of 4 ml total volume was carried out at 2 mJ energy per pulse and repetition rate of laser pulses of 10 kHz. Estimated diameter of the laser beam waist was 30 μm, which corresponds to laser fluence in the liquid up to 125 J/cm$^2$. Bright cylinder made of plasma appeared 2-3 mm above the window inner surface and looked continuous for eye.

Amperometric molecular hydrogen sensor was used to monitor the concentration of $H_2$ in the space above the liquid surface. Excessive pressure in the system was released to ambient air through a glass capillary dipped in 2 mm thick water layer. In this case the total pressure in the cell was equal to atmospheric one. Inner electrolyte of the sensor is separated from the cell atmosphere by a membrane pervious only to $H_2$. The sensor indicates either the concentration of H2 in μg/l (mg/l) or its partial pressure in Torrs. Calibration of the sensor was performed in air (no $H_2$) and in 1 atmosphere of $H_2$. The atmosphere pressure was checked for each day of measurements since is changes with time. The precision of $H_2$ concentration measurements is 5%. Total volume of the atmosphere above the water level can be estimated as 10 ml. The characteristic time of sensor response in this geometry is about 5 min.

Molecular oxygen sensor, also amperometric one, was attached to the atmosphere above the liquid. Initial content of $O_2$ was set to 20% at atmosphere pressure, and then the calibration was performed automatically according to sensor software. Both sensors allowed independently monitor the content of $O_2$ and $H_2$ in the atmosphere above the irradiated liquid during and after laser exposure. Moreover, both sensors were used for measurements of $H_2$ and $O_2$ content in the liquid right after laser exposure.

For measurements of hydrogen peroxide content we used a sensitive method for quantitative measurement of hydrogen peroxide employing enhanced chemiluminescence in a peroxidase–luminol–p-iodophenol system. The reaction was quantified with a Biotoks USE luminometer (Engineering Center "Ekologiia", Russia). The luminometer was used in single photon counting mode. Before measurements samples were placed in polypropylene vials (Beckman, USA) and 150 µl of 'count solution' containing $10^{-2}$ M Tris–HCl buffer, pH 8.5, $5 \times 10^{-5}$ M 4-iodophenol, $5 \times 10^{-5}$ M luminol and horseradish peroxidase ($10^{-9}$ M for nanomolar hydrogen peroxide measurement) was added [5]. The 'count solution' was prepared immediately prior to measurement. The initial $H_2O_2$ concentration used for calibration was measured spectrophotometrically at 240 nm taking the absorption molar coefficient value of 43.6 $M^{-1}cm^{-1}$ [6] into account. The sensitivity of the method allows one to evaluate $H_2O_2$ down to concentration of 1 nM [7].

Results and discussion

Hydrogen and oxygen yields dependence of exposure time and laser fluence in aqueous colloidal solution are presented in Fig. 1, a,b. The interesting phenomenon consists in the dependence of $O_2$ concentration on laser exposure time. $O_2$ concentration decreases at the beginning of laser exposure and only then starts to grow. On the contrary, $H_2$ concentration only increases with time of exposure. This difference is probably due to the fact that the initial $O_2$ concentration is around 20%. After the onset of laser exposure emitted $H_2$ pushes initial $O_2$ out of measuring volume. Then the $O_2$ concentration starts to increase due to emission of $O_2$ that originates from $H_2O$ dissociation.

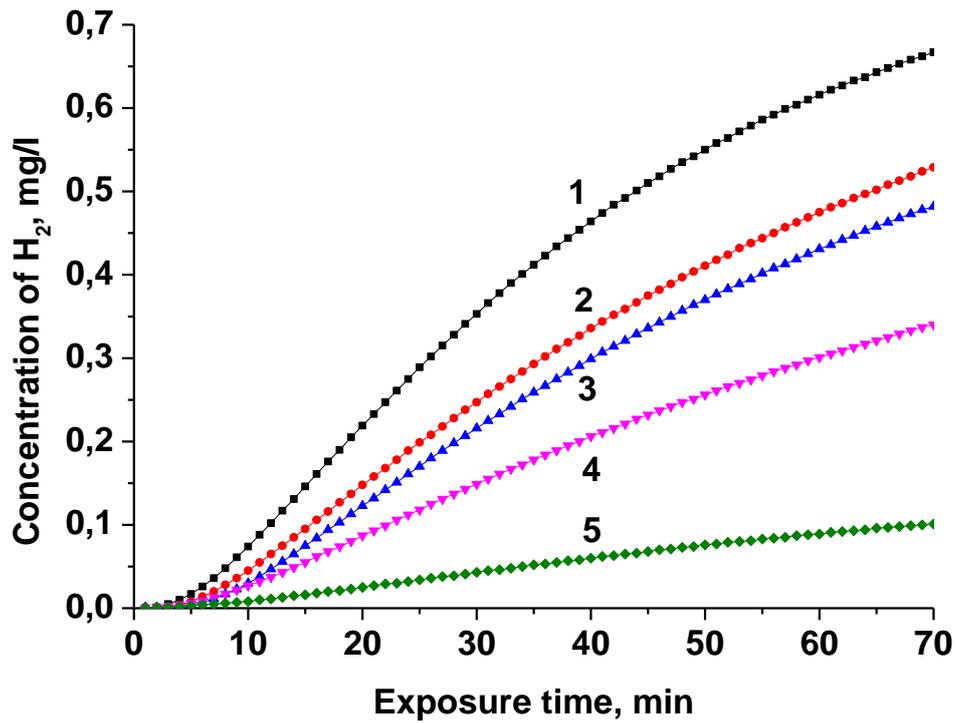

Fig. 1,a. Concentration of $H_2$ in atmosphere above the liquid as the function of time of laser exposure and laser fluence in the liquid, 1 -125, 2 – 124, 3 – 123, 4 – 118, and 5 – 110 J/cm$^2$.

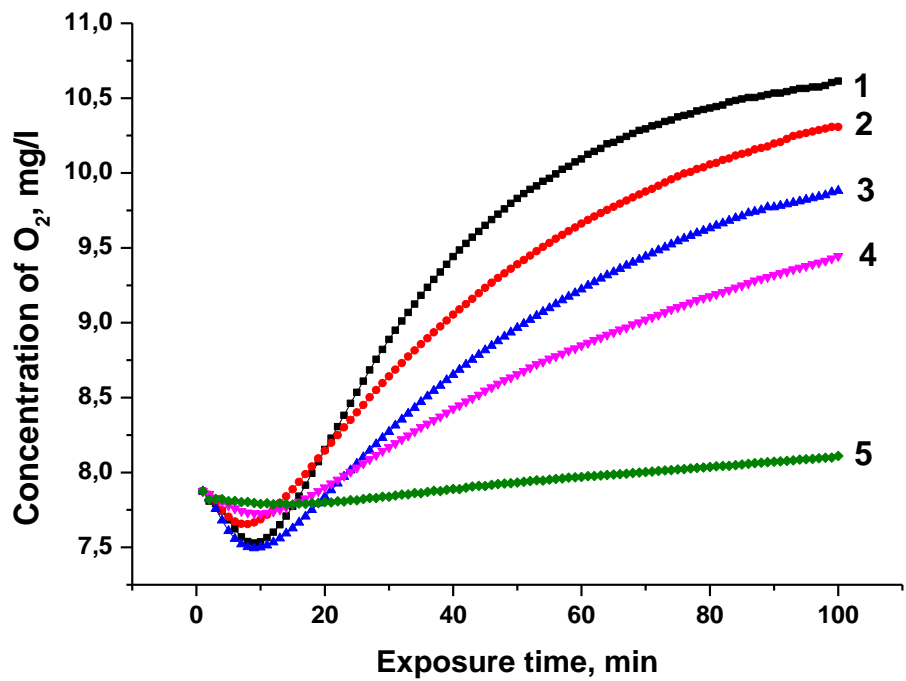

Fig. 1,b. Concentration of $O_2$ in atmosphere above the liquid as the function of time of laser exposure and laser fluence in the liquid, 1 -125, 2 – 124, 3 – 123, 4 – 118, and 5 – 110 J/cm$^2$.

The lower is the laser fluence, the more time is needed to reach the saturation level of $H_2$. In general, however, it was found that the temporal dependence of gas pressure on exposure time follows sigmoidal dependence (see Figs. 1 a, b). Then the $H_2$ content at relatively low laser fluence (<140 J/cm$^2$) was approximated by such a curve, and the pressure was taken to that reached at infinitely long time of laser exposure.

One can see that the relative content of $H_2$ shifts towards hydrogen with the increase of laser fluence. Indeed, there is no $H_2$ before the onset of laser exposure, while $O_2$ content corresponds to that one in ambient atmosphere (20%).

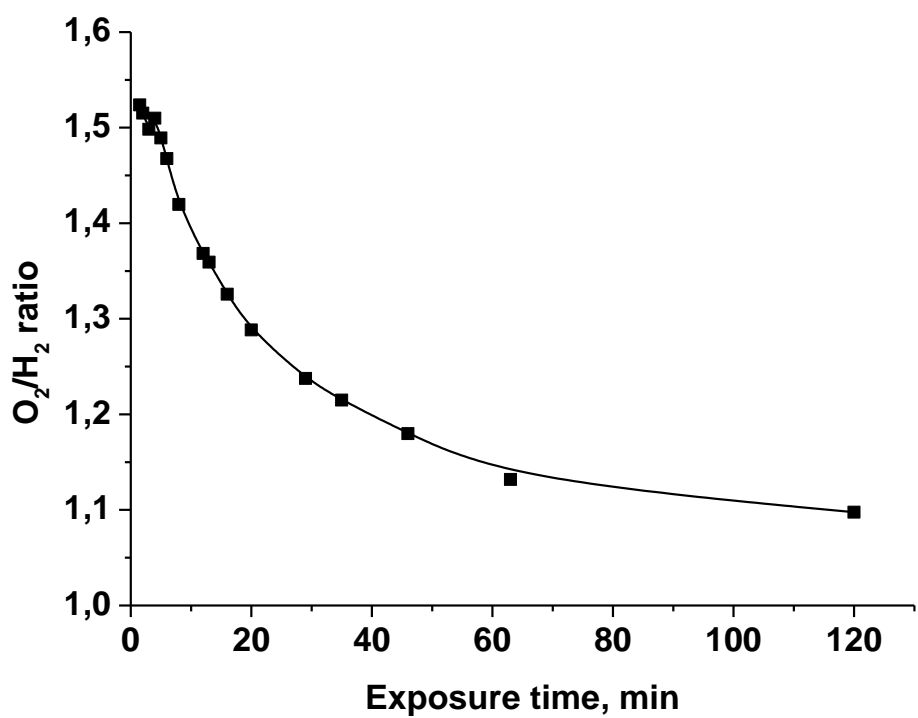

Fig. 2. Dependence of $O_2/H_2$ ratio in number of molecules on the exposure time. Laser fluence of 118 J/cm$^2$.

The balance between $H_2$ and $O_2$ also depends on the laser fluence inside the liquid at otherwise equal experimental conditions. Fig. 4 presents the dependence of this ratio on the laser fluence. At the threshold of plasma channel formation $H_2$ content and $H_2/O_2$ ratio are close to 0. At higher fluences $H_2/O_2$ ratio increases and reaches the plateau value of 1.4 at fluence above 130 J/cm$^2$. It should be stressed once again that the stationary levels of gases content at low laser fluence were obtained from sigmoidal approximation of the initial dependences.

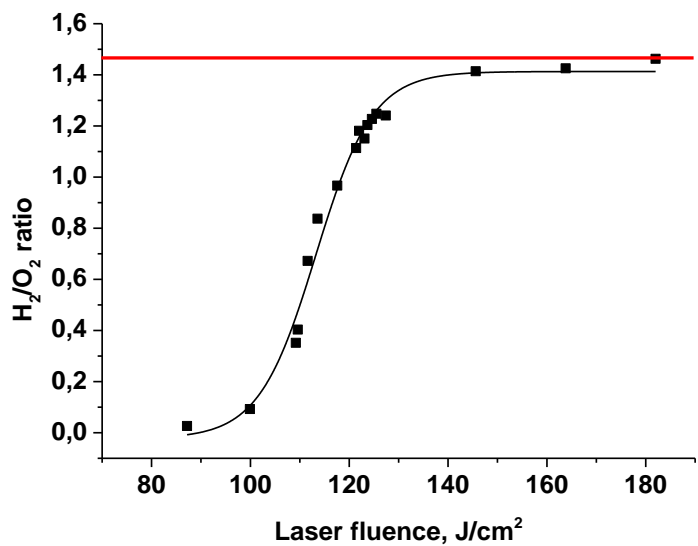

Fig. 3. Ratio $H_2/O_2$ in number of molecules as the function of laser fluence in the liquid.

The general reaction of $H_2O$ decomposition can be written as follows: $2H_2O = 2H_2 + O_2$. In this case $H_2/O_2$ ratio is 2. From Fig. 3 it can be concluded that the ratio of 1.4 means the deviation from stoichiometric decomposition. The lack of oxygen can be explained by formation of oxygen-containing products, such as $O_3$ or $H_2O_2$. Ozone $O_3$ is not stable while hydrogen peroxide is stable product. Indeed, the analysis of the colloidal solution subjected to laser exposure according to the analytical route described above shows the presence of $H_2O_2$. The concentration of hydrogen peroxide, unlike $H_2$ and $O_2$, linearly scales with laser exposure time of the colloid (Fig. 4).

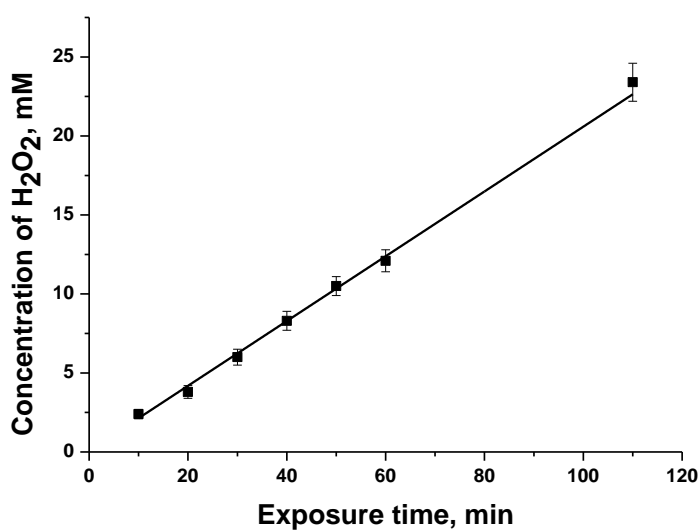

Fig. 4. Dependence of $H_2O_2$ concentration in mmole/l on the time of laser exposure at repetition rate of laser pulses of 10 kHz.

In another set of experiments an isolated tube with water was immersed into the same cell. The tube was protected from laser radiation by Al foil. It was found however that the formation of $H_2O_2$ also takes place in this isolated volume. The only possibility of formation of this stable product is the action of ultrasound produced by laser breakdown. The formation of hydrogen peroxide under sonication with ultrasound is well-known in literature [8, 9]. Our independent measurements performed with the help of piezoelectric sensor show that the acoustical spectrum of the sound inside the liquid is fairly wide reaching the frequencies of order of several MHz. For the given geometry of the cell there is also a possibility to produce standing ultrasound waves that are usually considered as a source of $H_2O_2$ formation. However, the concentration of $H_2O_2$ induced by ultrasound is typically 3 orders lower than that produced in the cell itself.

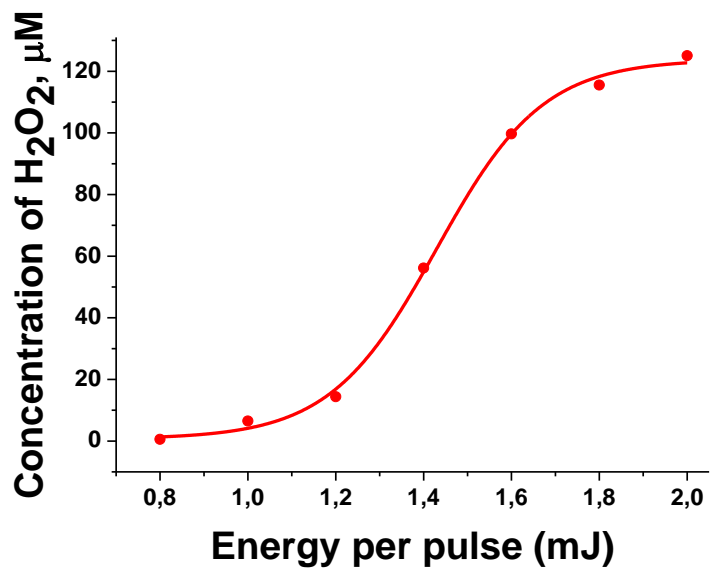

Fig. 5. Dependence of $H_2O_2$ concentration in μmole/l on the energy per laser pulse. Total exposure time of 5 minutes at repetition rate of laser pulses of 10 kHz. The size of symbols is equal to the error of measurements.

The formation of $H_2O_2$, as well as $H_2$ and $O_2$, strongly depends on the laser fluence inside the liquid (see Fig. 3 for gases). Qualitatively, the dependence of $H_2O_2$ yield is similar to that of gaseous products (Fig. 5).

The breakdown of the liquid occurs owing to the development of electron avalanche. The NPs serve as centers of initial electrons, and the yield of products, $H_2O_2$ in particular, should depend on the initial concentration of NPs in the colloidal solution. The dependence of $H_2O_2$ concentration on the concentration of Au NPs is presented in Fig. 6. Au NPs were generated by ablation of a bulk Au target in water. The initial size of Au NPs is 10 nm, as measured by a measuring disk centrifuge.

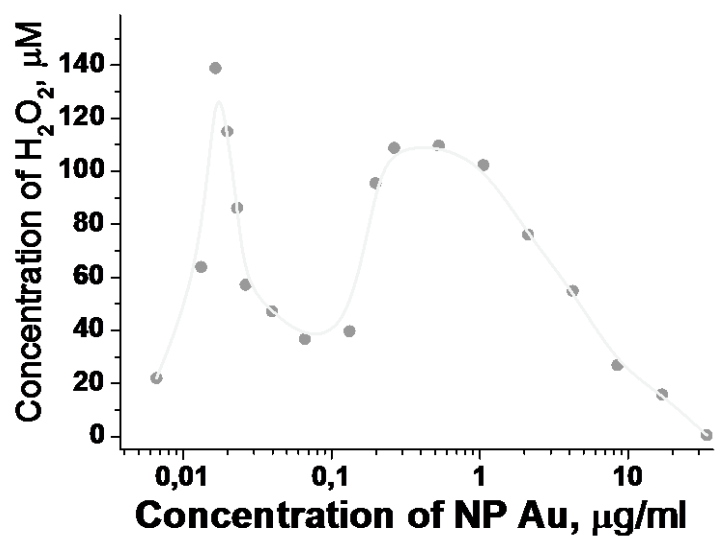

Fig. 6. Dependence of $H_2O_2$ production in μmole/l on the concentration of Au NPs. The size of symbols is equal to the error of measurements.

The first maximum in Fig. 6 could be assigned to the enhanced probability of laser liquid breakdown since Au NPs serve as a center of absorption of laser radiation. The second maximum in Fig. 6 may be due to the enhancements of the laser field on Au NPs owing to effect of plasmon resonance.

The formation of gaseous stable products ($H_2$, $O_2$) is interpreted as direct dissociation of water molecules by electron impact from breakdown plasma. Indeed, there are a number of channels that may lead to formation of these gases [4]. The electron energy required for these processes should be higher than 20 eV. This indirectly means that there are electrons in the breakdown plasma with such energy, though the energy of laser photons is only 1 eV. High energy of electrons can be assigned to the absorption of laser radiation by breakdown plasma. This absorption increases the electron temperature, which is finally sufficient for dissociation of water molecules. The formation of $H_2O_2$ in the present experimental conditions is less evident. Typically its formation is associated with radicals of OH. Combination of OH radicals with another radical or neutral OH may be the source of $H_2O_2$ formation during laser breakdown of the colloidal solution. Therefore, the formation of $H_2O_2$ proceeds via formation of unstable transient products.

Conclusion

Thus, formation of stable products of laser breakdown of aqueous colloidal solutions of nanoparticles has been experimentally studied. The ration $H_2/O_2$ of 1.4 is shifted to $H_2$ and depends on the laser fluence inside the liquid. Deviation of stoichiometry can be attributed to simultaneous formation of $H_2O_2$ in the same solution. Concentration of $H_2O_2$ increases linearly with the time of laser exposure of the solution. There are at least two sources of the formation of this product, namely, plasma of laser breakdown and ultrasound waves induced in the liquid during the breakdown. One may conclude that the investigated stable products accompany any process of laser ablation of solids in liquids.

Acknowledgements

The work was partially supported by Russian Foundation of Basic Researches, grants 15-02-04510_a, 16-02-01054_a RF President grant MK-3606.2017.2.